\documentclass[11pt,a4paper]{article}
\usepackage{amsmath,amsthm,amssymb}
\usepackage[utf8]{inputenc}
\usepackage[T1]{fontenc}
\usepackage{subfiles}
\usepackage{graphicx}
\usepackage{multirow}
\usepackage{caption}
\usepackage{booktabs}
\usepackage{enumitem}
\usepackage[obeyspaces]{url}
\usepackage{kotex}
\usepackage{lineno,hyperref}
\usepackage{indentfirst}
\usepackage{listings}
\usepackage{amsmath}
\usepackage{makecell}
\usepackage{array}
\usepackage{algorithm}
\usepackage{algpseudocode}
\usepackage{authblk}
\usepackage{url}
\usepackage{xurl}

\usepackage{xcolor}
\colorlet{kw}{blue}
\definecolor{com}{rgb}{0,0.6,0.3}

\algrenewcommand{\algorithmiccomment}[1]{\textcolor{kw}{\hfill$\triangleright$\texttt{#1}}}

\title{A Method for Decrypting Data Infected with Hive Ransomware}

  \author[a]{Giyoon Kim$^1$}
  \author[a]{Soram Kim$^1$}
  \author[a]{Soojin Kang$^1$}
  \author[a,b]{Jongsung Kim \thanks{Corresponding author \\ $^1$ These authors contributed equally to this work.
  \\ Email addresses : gi0412@kookmin.ac.kr (Giyoon Kim), kimsr2040@kookmin.ac.kr (Soram Kim), szin31@kookmin.ac.kr (Soojin Kang), jskim@kookmin.ac.kr (Jongsung Kim) \\
  URL: \url{https://dfnc.kookmin.ac.kr}}}

  \affil[a]{Dept. of Financial Information Security, Kookmin University,77 Jeongneung-Ro, Seongbuk-Gu, Seoul, 02707, Korea}
  \affil[b]{Dept. of Information Security, Cryptology, and Mathematics, Kookmin University,77 Jeongneung-Ro, Seongbuk-Gu, Seoul, 02707, Korea}

\begin{document}
	
\maketitle

\begin{abstract}

Among the many types of malicious codes, ransomware poses a major threat. 
Ransomware encrypts data and demands a ransom in exchange for decryption.
As data recovery is impossible if the encryption key is not obtained, some companies suffer from considerable damage, such as the payment of huge amounts of money or the loss of important data. 
In this paper, we analyzed Hive ransomware, which appeared in June 2021.
Hive ransomware has caused immense harm, leading the FBI to issue an alert about it.
To minimize the damage caused by Hive Ransomware and to help victims recover their files, we analyzed Hive Ransomware and studied recovery methods.
By analyzing the encryption process of Hive ransomware, we confirmed that vulnerabilities exist by using their own encryption algorithm.
We have recovered the master key for generating the file encryption key partially, to enable the decryption of data  encrypted by Hive ransomware.
We recovered 95\% of the master key without the attacker's RSA private key and decrypted the actual infected data.
To the best of our knowledge, this is the first successful attempt at decrypting Hive ransomware. 
It is expected that our method can be used to reduce the damage caused by Hive ransomware.

\end{abstract}


\section{Introduction}
The COVID-19 pandemic has increased the popularity of working from home and the use of online environments~\cite{mouratidis2021covid}.
Simultaneously, the number of cyber threats such as phishing, spam, Trojan horses, and ransomware targeting general users and businesses has increased. 
Newly discovered examples of ransomware have increased rapidly during the pandemic, and ransomware attacks and damage cases have increased correspondingly~\cite{SecurityMagazine}.

Ransomware operates systematically, encrypting data and demanding ransoms from victims in exchange for not distributing the stolen data and giving the data decryption method~\cite{richardson2017ransomware}.
The targets of attacks have also expanded, from general companies to companies that operate hospitals and major infrastructure, causing secondary damage to service users. 
Therefore, many companies under attack have no choice but to pay the ransom demanded. Ransomware attacks targeted several hospitals during the pandemic.
In February 2021, the Egregor ransomware attacked the Dax-Côte d'Argent Hospital Center, and the COVID-19 vaccination clinic needed to be closed~\cite{Hopitalde}.
In April and May 2021, health care providers in Australia and the United States, UnitingCare Queensland and Scripps Health, respectively, had problems with their operations due to ransomware attacks~\cite{UnitingCareQueensland,ScrippsHealth}.
In May 2021, Ireland's health service, Health Service Executive (HSE), was owing suspended due to an attack by the Conti ransomware~\cite{HSE}.
In 2021, a number of ransomware attacks also hit companies involved in industries such as transportation, food, and oil, which are the basis of modern life.
In April, Dutch transportation provider Bakker Logistiek was infected with ransomware, causing a cheese shortage~\cite{Dutchsupermarkets}.
In May, meat food giant JBS Foods stopped production facilities owing to an attack by REvil ransomware, and paid \$11 million to restore systems~\cite{JBS}.
In the same month, Colonial Pipeline, the largest fuel pipeline in the United States, was attacked by the DarkSide ransomware and had to shut down a 5,500-mile-long fuel pipeline~\cite{Colonial}.
This attack resulted in a regional emergency declaration in 17 states. Oil prices soared, and Colonial Pipeline eventually paid \$5 million.
Ransomware attacks continue to occur, and the amount of damage is gradually increasing.
To respond to this threat, countries around the world have announced plans and policies to respond to ransomware.
In 2021, several policies and warnings were announced related to various ransomware response, including Hive ransomware~\cite{homeaffairs,Canadian,weekend_alert,FBI_Hive}.
These policies are a precautionary measure against ransomware, and in the event of a ransomware infection, the use of data backed up by individuals and companies is suggested as a post-response measure.
However, without backup data, there is a limit to the recovery of data infected with ransomware.
Therefore, it is necessary to study ransomware to determine the possibility of decrypting encrypted data without using backup data.

The Hive ransomware analyzed in this paper first appeared around June 2021.
The attackers run an information leakage site called ``Hive Leak" to distribute data stolen from victims if they do not pay the ransom~\cite{hivebppage}.
Hive ransomware has been continually attacking targets since its appearance, resulting in damage to healthcare companies and other companies.
Altus Group, a data analysis and software provider in the commercial real estate industry, was infected with Hive ransomware and leaked data to Hive Leaks~\cite{Altus}.
Memorial Health System, an American medical organization, was attacked by Hive ransomware, leading to the cancellation of treatment and examinations, and some patient data were leaked to Hive Leaks~\cite{Memorial}.
The Missouri health center in the US and Macquarie Health Corporation in Australia were also attacked by Hive ransomware~\cite{Missouri,Macquarie}.
MediaMarkt, a European electronics outlet, was infected with Hive ransomware, causing IT systems and services to be suspended~\cite{MediaMarkt}.
As many such cases of damage have occurred, the U.S. Federal Bureau of Investigation (FBI) issued a Hive ransomware warning on August 27, 2021, two months after the Hive ransomware appeared~\cite{FBI_Hive}.

Recently, many ransomware attacks have been found to use a hybrid encryption scheme that encrypts user files with a symmetric cipher, and stores the encryption keys used with an asymmetric cipher. 
Most ransomware uses secure algorithms such as AES~\cite{rfc3826} and RSA~\cite{rfc8017} for encrypting files. 
Therefore, if an attacker’s private key is not obtained, it is difficult to decrypt the encrypted files.
However, certain ransomware may use a self-developed encryption algorithm when encrypting files. 
If attackers cryptographically misconfigure the ransomware, a cryptographic vulnerability can occur. 
The Hive ransomware encrypts a victim’s file using an encryption algorithm developed by the Hive programmers. 
We analyzed Hive ransomware and discovered the detailed operation process of Hive ransomware. 
Hive ransomware uses a hybrid encryption scheme, but uses its own symmetric cipher to encrypt files.
We were able to recover the master key for generating the file encryption key without the attacker’s private key, by using a cryptographic vulnerability identified through analysis. 
As a result of our experiments, encrypted files were successfully decrypted using the recovered master key based on our mechanism.
To the best of our knowledge, this is the first successful attempt at decrypting the Hive ransomware. 

We experimentally demonstrated that more than 95\% of the keys used for encryption could be recovered using the method we suggested. 
Our contributions are summarized as follows:

\begin{enumerate}

    \item We identified the way in which Hive ransomware generates and stores master key for victim files.
    Hive ransomware generates 10MiB of random data, and uses it as an master key.
    For each file to be encrypted, 1MiB and 1KiB of data are extracted from a specific offset of the master key and used as a keystream.
    The offset used at this time is stored in the encrypted file name of each file.
    Using the offset of the keystream stored in the filename, it is possible to extract the keystream used for encryption.
    
     \item We analyzed the Hive ransomware to uncover its operation process and a newly developed encryption algorithm process.
     Hive ransomware encrypts files by XORing the data with a random keystream that is different for each file.
     We found that this random keystream was sufficiently guessable.

    \item We suggested a method for decrypting encrypted files without the attacker’s private key. 
    We found that the Hive ransomware does not use all bytes of the master key encrypted with the public key. 
    Using our proposed method, more than 95\% of the master key used for generating the encryption keystream was recovered. 
    Most of the infected files could be recovered by using the recovered master key. 
    We present experimental results for the case of recovering files using our proposed method.

\end{enumerate}

\section{Related Work}
    Ransomware research is being conducted in proactive as well as reactive ways.
    Proactive ransomware research uses domain knowledge of ransomware to minimize ransomware damage, and has various focus areas such as detection, prevention and blocking, and analysis automation.
    
    Kharraz et al. analyzed how ransomware attacks evolved by analyzing 1,359 ransomware samples collected between 2006 and 2014~\cite{kharraz2015cutting}.
    They suggested that ransomware attacks could be prevented by tracking changes in the Master File Table (MFT) or the types of I/O Request Packets by investigating the file system activity in detail.
    Krzysztof et al. analyzed the CryptoWall ransomware, which communicates with the command and control (C\&C) server to receive the encryption key, collected the domain, URL and IP information used by the ransomware, and used this information to develop a system to block the infection by this ransomware~\cite{cabaj2018software}.
    Suhyeon Lee et al. suggested a method of changing the extensions of files targeted by ransomware to random extensions to avoid ransomware attacks.
    This approach prevents the ransomware from encrypting user data, which is the main target of the ransomware~\cite{lee2019ransomware}.
    Scaife et al. proposed an early warning system for ransomware attacks, archieved by detecting processes that manipulate user data and forcibly shutting down the system when such behaviors are found~\cite{scaife2016cryptolock}.
    To overcome the limitations of signature-based and static ransomware detection methods, Zhi-Guo et al. developed a model to detect ransomware by monitoring the API calls used by ransomware, and generating flow graphs based on these data~\cite{chen2017automatic}.
    Alhawi et al. analyzed the network traffic generated when ransomware communicates with a C\&C server to detect ransomware using network traffic features~\cite{alhawi2018leveraging}.
    Mehnaz et al. proposed a method for detecting ransomware processes in real time using a decoy file~\cite{mehnaz2018rwguard}.
    They distinguished ransomware encryption from benign file changes by monitoring process and file changes, and hooking in Microsoft's CryptoAPI.
    To overcome the limitations of existing ransomware detection, Kok et al. developed a pre-encryption algorithm that combines a signature repository that identifies signatures matching ransomware and a learning algorithm that can detect known as well as unknown crypto ransomware~\cite{kok2020evaluation}.
    Continella et al. suggested a ransomware detection method based on I/O monitoring, the ratio and type of files with changed file names, and the frequency of folder listing operations~\cite{continella2016shieldfs}.
    Sang-moon et al. developed a ransomware detection and response system by constructing a ransomware detection model by monitoring the entropy of encrypted files~\cite{jung2018ransomware}.
    Cabaj et al. proposed a countermeasure against ransomware that interferes with the connection to the C\&C server using Software-Defined Networking (SDN)~\cite{cabaj2016using}.
    Kharaz et al. proposed UNVEIL, a dynamic analysis system that detects ransomware by tracking changes in users' files or desktops~\cite{kharaz2016unveil}.
    Xu et al. proposed CryptoHunt, a tool to detect cryptographic functions such as TEA, AES, RC4, MD5, and RSA in obfuscated binary files~\cite{xu2017cryptographic}.
    Li et al. developed K-Hunt, a tool to find and identify encryption keys by tracing how encryption keys are generated and propagated in binary executable files~\cite{li2018k}.
    Hill et al. developed CryptoKnight, which classifies cryptographic primitives in compiled binary executables using deep learning, to identify the algorithms used in crypto ransomware~\cite{hill2018cryptoknight}.
    Villalba et al. developed a tool that captures the pop-up windows displayed by the ransomware, and uses optical character recognition technology to analyze the messages and dump the memory, allowing analysts to analyze the ransomware~\cite{villalba2018ransomware}.

    Similar to our research, there are also researches on decryption methods for files infected with ransomware.
    In this case, research is generally conducted by targeting specific ransomware.
    Sehoon et al. analyzed the Magniber v2 ransomware and developed a method to decrypt Magniber v2 without the attacker's private key~\cite{lee2021magniber}.
    They reduced cryptographic key candidates by using the vulnerability of the Pseudo Random Number Generator (PRNG) that Magniber v2 ransomware uses to generate the cryptographic key, and conducted a study to find the cryptographic key using the padding verification and statistical randomness test of NIST SP800-22~\cite{rukhin2001nist}.
    Sehoon et al. analyzed the encryption process for major ransomware and five new ransomware applications (Gandcrab v5, Clop, Sodinokibi, Phobos, and LooCipher) in 2019, and presented scenarios that can be decrypted~\cite{lee2019study}.
    They showed that Gandcrab can be decrypted by leaking the attacker's public key, and LooCipher is decrypted by reproducing a weak random number generator.
    Kyungroul et al. suggested a method of recovering an original file from the backup system by detecting the ransomware-infected files synchronized with the backup system~\cite{lee2019machine}.
    They used entropy to analyze the characteristics of encrypted files, and machine learning algorithms to classify infected files based on this information.
    Kolodenker et al. produced PayBreak, which hooks and stores the session keys generated for encrypting victim files in crypto ransomware. Victims can use the keys obtained by PayBreak to recover infected files~\cite{kolodenker2017paybreak}.
    Cheng et al. developed dptCry, a tool to find key information to decrypt when infected by crypto ransomware such as WannaCry is infected~\cite{cheng2019dptcry}.
    dptCry intercepts sensitive encryption key information using API hooking methods to monitor running processes in real time.
    Each study found a vulnerability in the software structure of the ransomware or found a cryptographic vulnerability to decrypt the data.
    In this respect, the above researches are similar to ours, but cannot be applied to the Hive ransomware analyzed in this paper.

\section{Hive ransomware analysis}
\label{hive_analysis}
The hash values of the Hive ransomware in our experiments are listed in Table~\ref{tab:sample_properties}. We analyzed the sample files with IDA pro v7.6 in VMware because Hive ransomware is written in the Go language. 
Each Hive ransomware sample is packed with UPX, and there is no anti-analysis element.

\begin{table}[ht]
\centering
\caption{Hash value of Hive ransomware sample files}
\label{tab:sample_properties}
\resizebox{0.9\textwidth}{!}{

\begin{tabular}{cl}
\toprule
Hash Type & \multicolumn{1}{c}{Hash Value} \\ \midrule
\multirow{5}{*}{SHA256} & 2F26EA19A8FDB167B8593E8EEC03C37248B6E5008F0B9EE5FB7D326CBE6500BF \\
 & 50AD0E6E9DC72D10579C20BB436F09EEAA7BFDBCB5747A2590AF667823E85609 \\
 & 88F7544A29A2CEB175A135D9FA221CBFD3E8C71F32DD6B09399717F85EA9AFD1 \\
 & 612E5FFD09CA30CA9488D802594EFB5D41C360F7A439DF4AE09B14BCE45575EC \\
 & E1A7DDBF735D5C1CB9097D7614840C00E5C4D5107FA687C0AB2A2EC8948EF84E \\
 \bottomrule
\end{tabular}

}
\end{table}

\subsection{Encryption process}

\begin{figure}[h]
    \centering
    \includegraphics[width=0.9\textwidth]{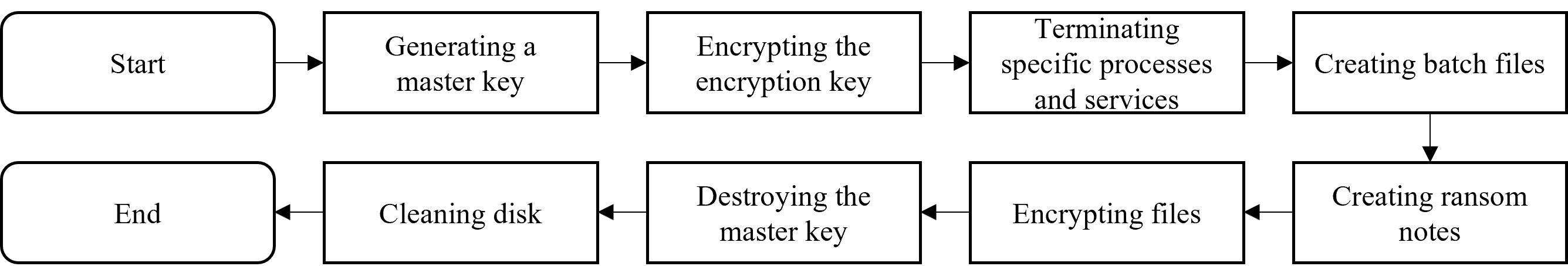}
    \caption{Entire encryption process of Hive ransomware}
    \label{fig:entire_process}
\end{figure}

The entire encryption process is described in Fig.~\ref{fig:entire_process}.

    \paragraph{\textbf{Generating a master key}}

    The Hive ransomware generates a 0xA00000 bytes (10MiB) random data using the rand function of the crypto package in the Go language. 
    The generated 10MiB data is the master key of Hive ransomware, and is used to generate a keystream for data encryption (details in Section~\ref{sec:encryption_process}).
    
    \paragraph{\textbf{Encrypting the master key}}
    
    The Hive ransomware encrypts the random number using an RSA-2048-OAEP public key fixed in a ransomware executable file, and stores the encrypted master key file.
    If the Hive ransomware has administrator privileges, it is stored in \path{C:\\}; if not, it is stored in \path{C:\Users\<User_name>\AppData\Local\VirtualStore}. 
    The filename of the encrypted master key is \textit{base64url\_encoded\_string.key.hive}. 
    The encoded string is the value encoded by hashing the encrypted master key with the MD5 hash function.

    \paragraph{\textbf{Terminating specific processes and services}}
    
    The Hive ransomware terminates specific processes and services as listed in Table ~\ref{tab:termination_list}.
    
    \begin{table}[h]
        \centering
        \caption{Termination list of processes and services}
        \label{tab:termination_list}
        \resizebox{0.7\textwidth}{!}{%
        \begin{tabular}{cc}
        \toprule
        Process          & Service                                              \\ \midrule
        mspub, msdesktop & bmr, sql, oracle, postgres, redis, vss, backup, sstp\\
        \bottomrule
        \end{tabular}%
        }
    \end{table}
    
    \paragraph{\textbf{Creating batch files}}

\begin{figure}[h]
        \centering
        \includegraphics[width=0.9\textwidth]{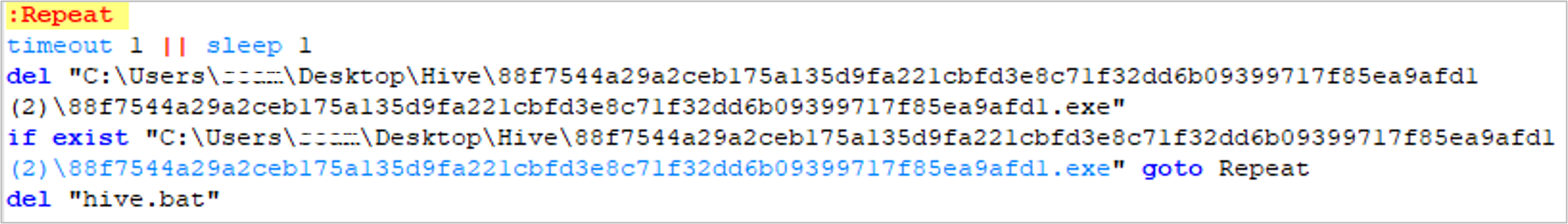}
        \caption{A hive.bat file of Hive ransomware}
        \label{fig:hive_bat}
    \end{figure}

\begin{figure}[h]
        \centering
        \includegraphics[width=0.5\textwidth]{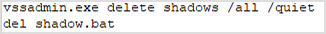}
        \caption{A shadow.bat file of Hive ransomware}
        \label{fig:shadow_bat}
    \end{figure}

    The Hive ransomware creates two batch files: hive.bat and shadow.bat. A hive.bat file deletes the Hive ransomware executable file, and then removes itself as well: commands are described in Fig.~\ref{fig:hive_bat} .
    A shadow.bat file deletes volume shadow copy (VSC) files, and then removes itself; the commands are described in Fig.~\ref{fig:shadow_bat}.
    
    \paragraph{\textbf{Creating ransom notes}}

\begin{figure}[h]
        \centering
        \includegraphics[width=1\textwidth]{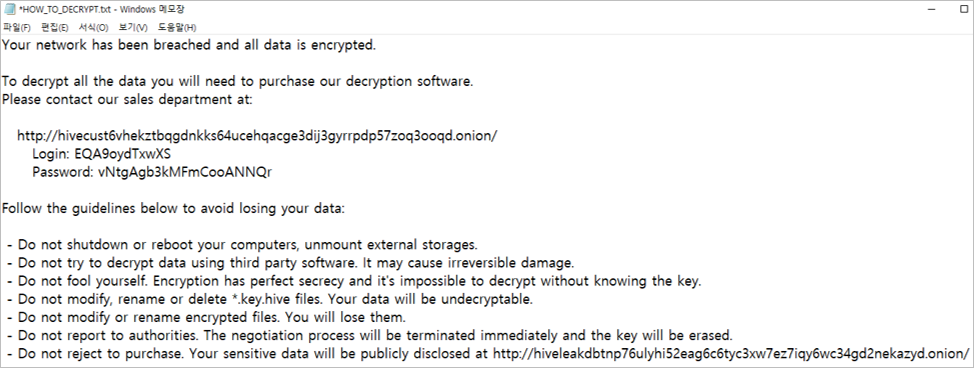}
        \caption{A Hive ransomware ransom note}
        \label{fig:ransomnote}
    \end{figure}

    The Hive ransomware creates a ransom note in each directory with the file name  \textit{HOW\_TO\_DECRYPT.txt}. The ransom note contains a warning; the URL and login information for decryption; and the guidelines to decrypt the user data, as described in Fig.~\ref{fig:ransomnote}.
    
    \paragraph{\textbf{Encrypting files}}

    The Hive ransomware encrypts all files except for .lnk files, executables, and files stored in \path{C:\Users\Windows} via an encryption process with ten threads in parallel. 
    However, \path{C:\Users\Program Files(x86)}, \path{C:\Users\Program Files}, and \path{C:\Users\ProgramData} paths are exceptions. 
    If the Hive ransomware gains administrator privileges, files in \path{C:\Users\Program Files(x86)}, \path{C:\Users\Program Files}, and \path{C:\Users\ProgramData} are encrypted. 
    If not, the files in \path{C:\Users\Program Files(x86)}, \path{C:\Users\Program Files}, and the \path{C:\Users\ProgramData} path are stored in \path{C:\Users\<User_name>\AppData\Local\VirtualStore} and encrypted. 
    In this case, the encrypted file and the original file coexist on the computer.
    The detailed file encryption process is described in Section~\ref{sec:encryption_process}.
    
    \paragraph{\textbf{Destroying the master key}}

    After encrypting the file, Hive Ransomware deallocates a randomly generated 10 MiB master key used to encrypt the file.
    Therefore, if we can take a memory snapshot after Hive Ransomware infection starts and before file encryption is finished, it is possible to recover the infected file when the master key existing in the memory is extracted.
    However, this is not easy as Hive ransomware infects very quickly.

    \paragraph{\textbf{Cleaning disk}}
    After infection, Hive Ransomware creates and deletes meaningless data in \path{C:\\} or \path{C:\Users\<User_name>\AppData\Local\VirtualStore} until the hard disk is full.This makes it impossible to restore encrypted files using residual data remaining on the disk.

\subsection{File encryption process}
\label{sec:encryption_process}

    Hive ransomware generates a data encryption keystream ($EKS$) that appears random for each file, and encrypts the file by XORing $EKS$ with the file.
    However, $EKS$ is created using two keystreams extracted from the previously created master key
    During the encryption process, only the part of the file, not the entire area, is encrypted.
    In this section, we describe details of the encryption process of Hive Ransomware.

    \paragraph {\textbf{Extracting two keystreams from the master key}}
    
    For each file encryption process, two keystreams from the master key are needed.
    Two keystreams are created by selecting two random offsets from the master key and extracting 0x100000 bytes (1MiB) and 0x400 bytes (1KiB) from the selected offset, respectively.
    The offsets are calculated as shown below, using two random numbers (R1, R2) of eight bytes each from the rand function of the Math package. 
    
    \begin{itemize}
        \item Keystream1 offset (SP1) : \texttt{R1 $\%$ 0x900000}
        \item Keystream2 offset (SP2) : \texttt{R2 $\%$ 0x9FFC00}
    \end{itemize}
    
\begin{figure}[!ht]
        \centering
        \includegraphics[width=0.85\textwidth]{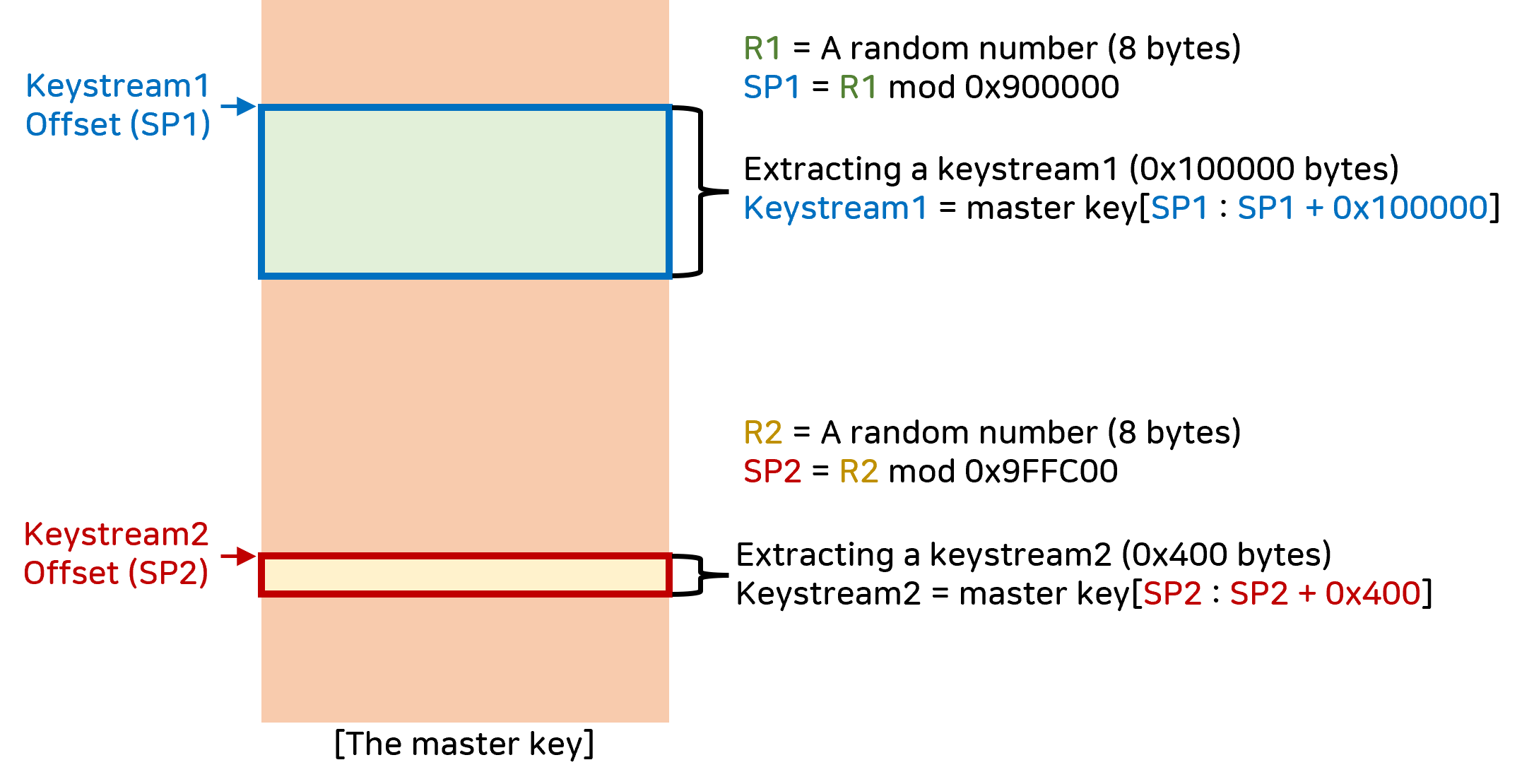}
        \caption{Selecting two keystreams from the master key}
        \label{fig:select_keystream}
    \end{figure}

    After selecting the keystream offsets, the two keystreams are extracted as specified in Fig.~\ref{fig:select_keystream}.
    The keystreams are used continuously until one file encryption process is finished.
    When the file is encrypted, R1 and R2 are stored in \textit{filename} in the little-endian format after encoding with base64url.
    The file name is generated using the following rule: Original Filename.\textit{base64url(MD5(Encrypted\_master\_key)$\parallel$R1$\parallel$R2)}.hive

    \paragraph {\textbf{Encrypting a file}}
    
    File encryption is performed using an XOR operation as follows:
    \begin{itemize}
        \item Encrypted data[i]$ \leftarrow$ Data[i] $\bigoplus$ Keystream1[i\%0x100000] $\bigoplus$ Keystream2[i\%0x400]
    \end{itemize}
    
    Given that Keystream1 and Keystream2 are used repeatedly by a modular operation, we can express the above operation as follows:
    
    \begin{itemize}
        \item $EKS$[i]$\leftarrow$ Keystream1[i] $\bigoplus$ Keystream2[i\%0x400] (i$\leftarrow$ 0,1,$\cdots$,0xFFFFF)
        \item Encrypted data$\leftarrow$ Data[offset] $\bigoplus$ $EKS$[offset\%0x100000]
    \end{itemize}
    At this time, the offset starts at 0 and increases by 1 sequentially, but it is not continuous.
    Offset increases by a total of 0x1000 times and then jumps by a certain amount.
    The non-encrypted data block size ($NBS$) that the offset jumps by depends on the size of the file to be encrypted and is calculated as presented in Table~\ref{tab:calculating_BS}.

    \begin{table}[h]
    \centering
    \caption{Calculating non-encrypted data block size based on a file size}
    \label{tab:calculating_BS}
    \resizebox{0.7\textwidth}{!}{%
    \begin{tabular}{lclc}\toprule
    \makecell[c]{File size (FS)}    & Temporary value (T)  & \makecell[c]{Non-encrypted data block size (NBS)} & The number of encrypted data blocks \\\midrule
    \makecell[c]{Not more than 4KiB\\(0x1000 bytes)} & - & \makecell[c]{0} & 1 \\[2ex]
    \makecell[c]{Under 128KiB \\ (0x20000 bytes)} & $T=(FS\gg12)$ &  & 1—31 (Avg. 16)\\[2ex]
    \makecell[c]{Under 1MiB\\(0x100000 bytes)} & $T=\frac{((FS\gg12)\times30)}{100}$ &  & 8—75 (Avg. 42)\\[2ex]
    \makecell[c]{Under 10MiB \\(0xA00000 bytes)} & $T=\frac{((FS\gg12)\times20)}{100}$ &  \makecell[c]{$NBS=\frac{FS-(T\ll12)}{T-1}$ \\[1.5ex] (if T==1, NBS=0)} & 50—510 (Avg. 281) \\[2ex]
    \makecell[c]{Under 100MiB\\(0x6400000 bytes)} & $T=\frac{((FS\gg12)\times10)}{100}$ &  & 255—2558 (Avg. 1406)\\[2ex]
    \makecell[c]{Under 1GiB\\(0x40000000 bytes)} & $T=\frac{((FS\gg12)\times5)}{100}$ &   & 1279—13106 (Avg. 1792)  \\[2ex]
    \makecell[c]{More than 1GiB\\(0x40000000 bytes)} & $T=\frac{((FS\gg12)\times1)}{100}$ &  & 2620$\leq$ \\ \bottomrule
    \end{tabular}%
    }
    \end{table}

\begin{figure}[!ht]
        \centering
        \includegraphics[width=0.9\textwidth]{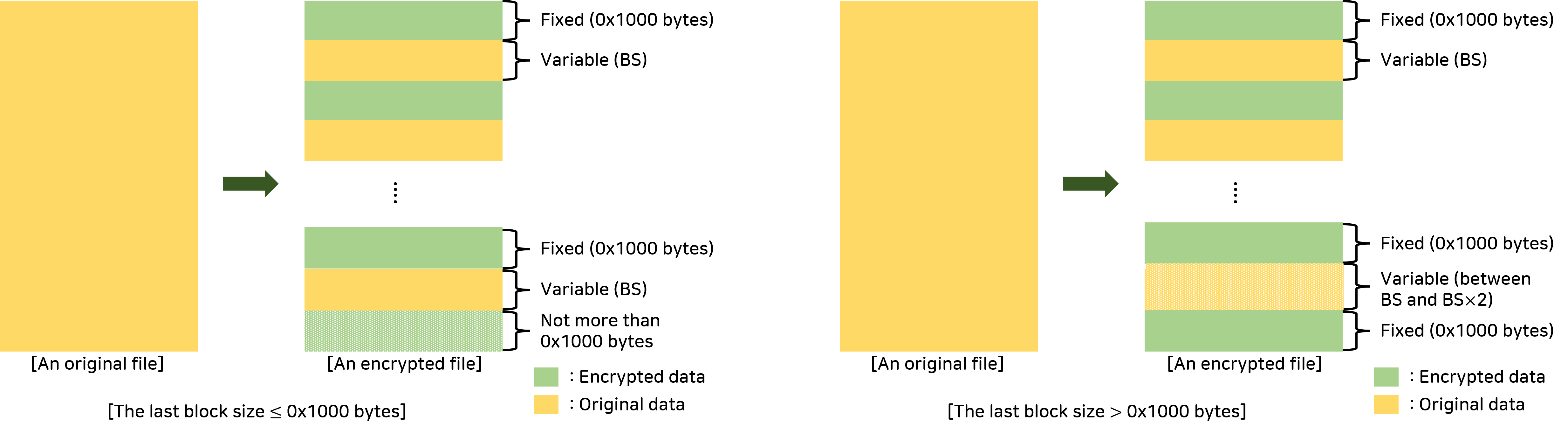}
        \caption{Encrypted file structure}
        \label{fig:encrypted_file_p3}
    \end{figure}

    The encrypted data block (0x1000 bytes) and the non-encrypted data block ($NBS$ bytes) appear alternately, as depicted in Fig.~\ref{fig:encrypted_file_p3}. However, in both, the last block size is variable. If the last block size is more than 0x1000 bytes, 0x1000 bytes of the last block are encrypted from the end of the file. If not, all bytes of the last block are encrypted.
    
     \paragraph {\textbf{Example of Hive ransomware infection}}
    Let us assume that the `test.jpg' file of size 0x667926 is infected with Hive Ransomware and the value of MD5(Encrypted\_master\_key) is 3636a6ca269b243753ff929c834d53ca.
    At this time, if R1 and R2 are randomly selected as 0x2e345b0798667926, 0x14199382ec72ddb6, then SP1 and SP2 become 0x667926 and 0x24f5b6.
    $EKS$ is created with two keystreams (master key[0x667926:0x767926], master key[0x24f5b6:0x24f9b6]).
    By the Table~\ref{tab:calculating_BS}, $NBS$ becomes 0x24f5b6, so 0x1000 bytes encryption and 0x24f5b6 bytes non-encrypted are repeated.
    As the size of last block exceeds 0x1000, `test.jpg' is encrypted in the form on the right part of Fig.~\ref{fig:encrypted_file_p3}.
    After the file data encryption is completed, the file name is changed to end the infection of `test.jpg'.
    According to the filename rule (\textit{base64url(MD5(Encrypted\_master\_key)$\parallel$R1$\parallel$R2)}), the infected file name is `text.jpg.NjamyiabJDdT\_5Kcg01TyiZ5ZpgHWzQutt1y7IKTGRQ.hive'.

\section{Hive ransomware decryption methodology}
\label{decryption_methodology}
In this section, we describe a file decryption method for the Hive ransomware using a cryptographic vulnerability. 
The Hive ransomware extracts two keystreams for file encryption from the master key, which is generated once at the beginning of each file encryption. The two keystreams are used to generate an $EKS$, and $EKS$ encrypts the data using XOR.
$EKS$ looks random, but the keystream to generate $EKS$ is partially reused when encrypting various files.
The encryption algorithm is an XOR operation, and the algorithm that generates $EKS$ is also XOR; therefore, it becomes easy to guess the keystreams.
We obtained equations consisting of XOR operations from encrypted files, and found the master key by solving the equations.
Our method to obtain the equations requires one of the following conditions:

\begin{itemize}
    \item Some of the original files corresponding to encrypted files should be available.
    \item There should be several encrypted files with known signatures, such as .pdf, .xlsx, or .hwp.
\end{itemize}

The main ways in which the original files corresponding to encrypted files can be obtained are described below. 
Unlike other ransomware, the Hive ransomware encrypts the Program files, Program files (x86), and ProgramData directories, which commonly store software files (Java, Python, Microsoft Office, and others) that are not related to the operating system. 
Therefore, we could easily obtain the original unencrypted files, as these software installation files can be obtained on the internet. 
Backup, synchronizing, and downloading files, and email attachments could also be utilized. By XORing the original and the encrypted file we can obtain the $EKS$.

It is also possible to determine the $EKS$ without the original unencrypted files. In this case, we use known file signatures. However, file signatures are short in comparison with the file data, therefore, many files with known signatures are required.

\subsection{Method for restoring the Hive ransomware master key}

The method described in this section aims to restore the master key by collecting as many as possible fragmented $EKS$ satisfying at least one of the two conditions.
If either of the two conditions is satisfied, the $EKS$ can be collected and the master key recovered.
 The core encryption algorithm of the Hive ransomware analyzed in Section~\ref{hive_analysis} is as follows:

 \begin{itemize}
    \item SP1$\leftarrow$ Keystream1 start offset (\texttt{R1 $\%$ 0x900000})
    \item SP2$\leftarrow$ Keystream2 start offset (\texttt{R2 $\%$ 0x9FFC00})
    \item Keystream1$\leftarrow$ Encryption\_key[SP1 : SP1+0x100000]
    \item Keystream2$\leftarrow$ Encryption\_key[SP2 : SP2+0x400]
    \item $EKS$[i]$\leftarrow$ Keystream1[i] $\bigoplus$ Keystream2[i\%0x400] (i$\leftarrow$ 0,1,$\cdots$,0xFFFFF)
    \item Encrypted data$\leftarrow$ Data[offset] $\bigoplus$ $EKS$[offset\%0x100000]
\end{itemize} 

\begin{figure}[!ht]
        \centering
        \includegraphics[width=0.8\textwidth]{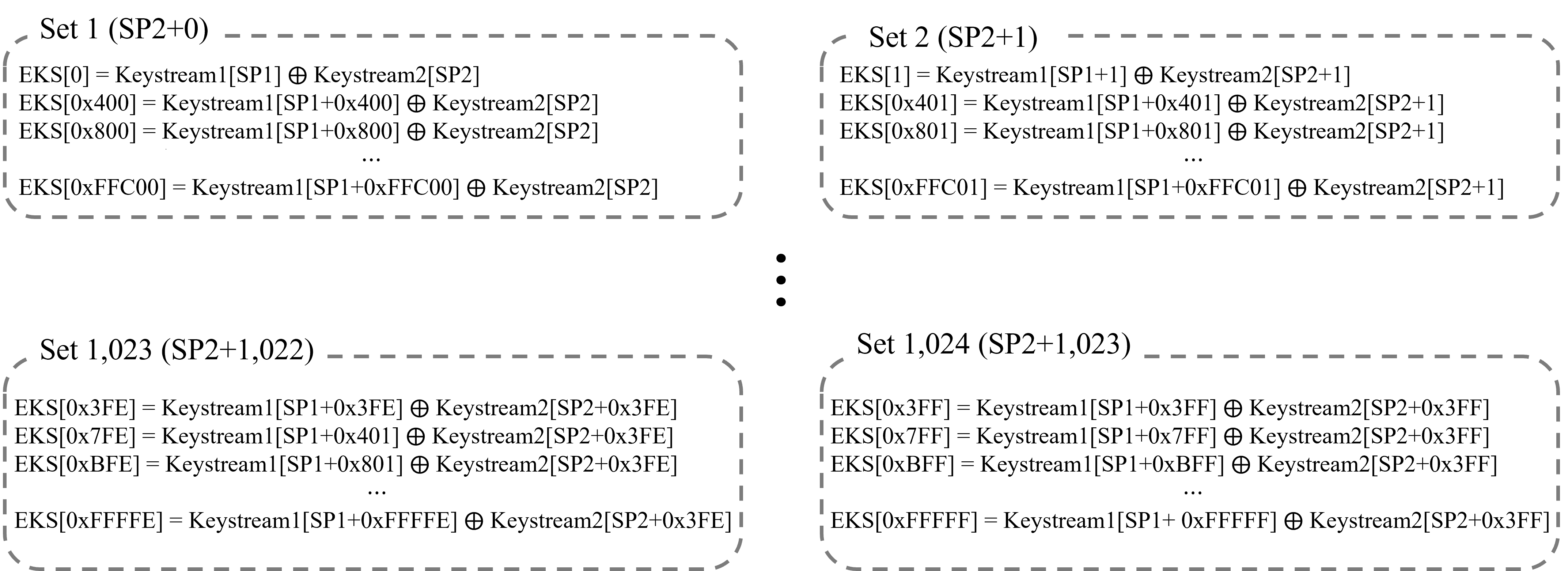}
        \caption{Process of the set separation of equations obtained from $EKS$}
        \label{fig:example}
\end{figure}

As can be seen from the core encryption algorithm, the $EKS$ is generated by repeating Keystream2, concatenating it 1,024 times, and then XORing it with Keystream1. 
The $EKS$ is generated as two keystreams, but these are data of a specific offset of the master key.
To be able to acquire up to 1MiB of continuous $EKS$ using the original file and the infected file is the same as being able to obtain 0x100000 XOR equations consisting of 0x100000 and 0x400 offsets of the master key.
As Keystream2 is used repeatedly, if the set is divided based on the repeated offset, the 0x100000 XOR relation is divided into 1,024 sets of size 1,024 (Fig.~\ref{fig:example}).

\begin{figure}[!ht]
        \centering
        \includegraphics[width=0.6\textwidth]{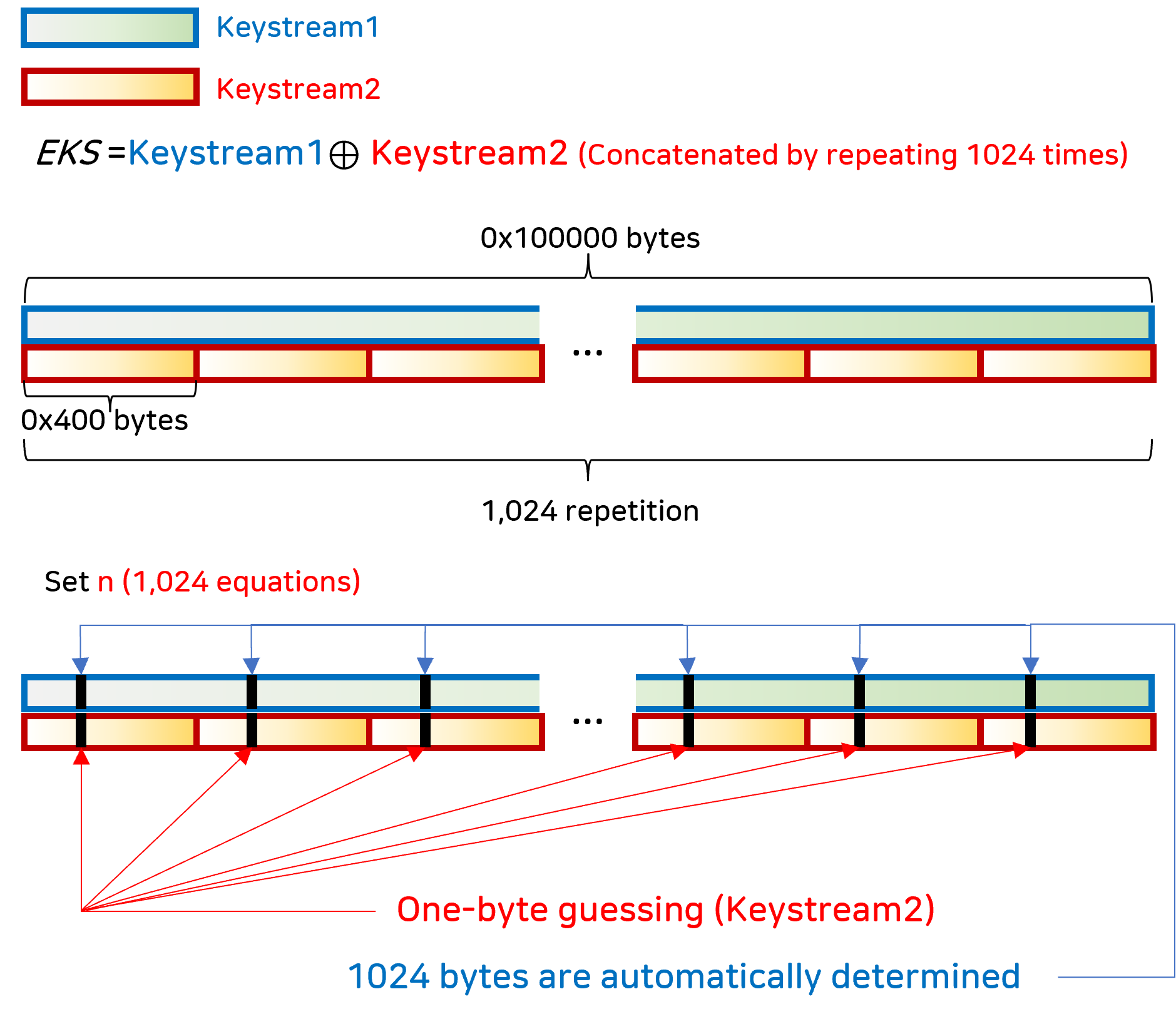}
        \caption{Process by which one byte is guessed to determine another byte (single file used)}
        \label{fig:part4_1byte}
    \end{figure}

As each set forms one simultaneous equation, if one byte of Keystream2 is guessed, 0x400 values are also determined by solving the equation. That is, if one $EKS$ is used, a maximum value of 1,025 bytes of the master key can be determined by guessing one byte (Fig.~\ref{fig:part4_1byte}).
    
1,025 bytes are determined depending on one byte; thus 256 cases are needed to find the actual master key. However, if two bytes are selected from among 1,025 bytes to generate the $EKS$, the byte guessed by XOR is removed, and using this approach, we can generated the $EKS$s that are uncollected.

\begin{itemize}
    \item Keystream1[SP1 + 1] $\bigoplus$ Keystream2[SP2 + 1] = 0x80 
    
    $\Longleftrightarrow$ Keystream1[SP1 + 1] = Keystream2[SP2 + 1] $\bigoplus$ 0x80 
    \item Keystream1[SP1 + 0x401] $\bigoplus$ Keystream2[SP2 + 1] = 0x88
    
    $\Longleftrightarrow$ Keystream1[SP1 + 1] = Keystream2[SP2 + 1] $\bigoplus$ 0x88 
    \item Keystream1[SP1 + 1] $\bigoplus$ Keystream1[SP1 + 0x401] = 0x08
\end{itemize}

The amount of data that can be guessed from one set of 1MiB $EKS$ is 1,025 bytes. 
Now consider the case of two sets.
When generating an $EKS$ by selecting each value from two independent sets, the deletion of the guess value is not performed with probability 1.
For example, suppose that two independent equations $x_1 \bigoplus y_1 = $0x80$, x_2 \bigoplus y_2 = $0x88 are obtained through $EKS$.
And suppose that the actual value was $x_1 = $0xFF$, y_1 = $0x7F$, x_2 = $0x23$, y_2 = $0xAB.
If we want to know the value of $x_1 \bigoplus x_2$, we have to know the correct $y_1 \bigoplus y_2$ value. 
Failure to find the correct $y_1 \bigoplus y_2$ results in the equation becoming infeasible.
In this case, we have to search 256 times to generate an equation that cannot be generated from the $EKS$.
When the correct equation is found, the two sets are chained to form one large set.
Finally, it takes $(n-1)\times256$ enumerations to form a system of equations from an independent set of size $n$.
As a result of our experiment, the number of independent sets $n$ was 1 when a large number of $EKS$ were collected. It was possible to recover a sufficient number of master key, even using one of the largest sets.

\begin{figure}[!ht]
        \centering
        \includegraphics[width=1\textwidth]{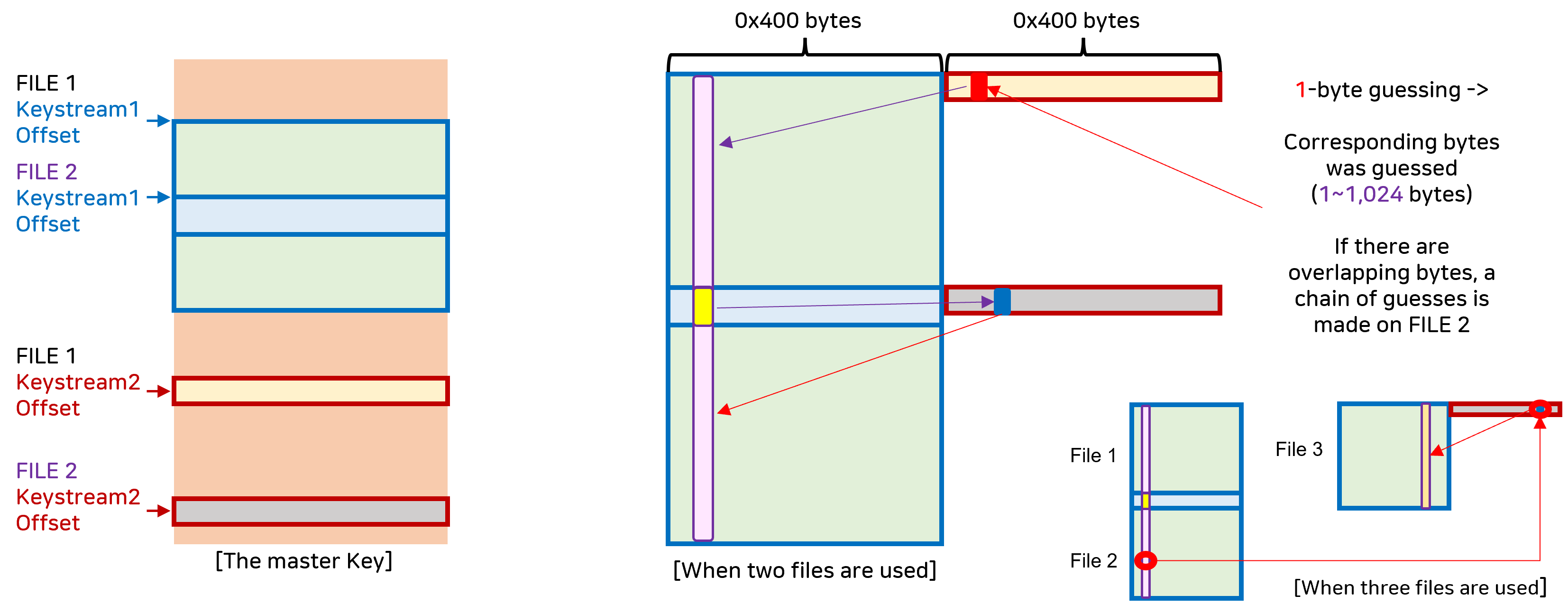}
        \caption{The process by which one byte is guessed to determine another byte (multiple files used)}
        \label{fig:part4_23byte}
    \end{figure}

Next, consider the case in which the $EKS$ is obtained from two encrypted files.
At this time, consider the case where the offset overlaps at least once among Keystreams for generating $EKS_1$ and $EKS_2$ (Fig.~\ref{fig:part4_23byte}). 
This problem is equivalent to finding an equation for chaining two sets independent of each other.
Therefore, if one offset overlaps, a maximum of 2,049 bytes of the master key can be found by guessing the overlapped offset; with this, the more $EKS$ obtained, the more the number of master key recovered at once is naturally accepted.

The following is the algorithmicization of the key recovery process based on the above process (Algorithms 1–3).

\newpage

\begin{algorithm}[!ht]
\caption{Calculation of the non-encrypted data block size}\label{alg:calc_BS}
\begin{algorithmic}
\Require File\_Size
\Ensure non-encrypted data block size $NBS$
\State \texttt{FS $\gets$ File\_Size}
\State \texttt{NBS $\gets$ 0}
\If{\texttt{FS $\leq$ 0x1000}}
    \Return \texttt{NBS}
\ElsIf{\texttt{FS < 0x20000}}
    \State \texttt{R$\gets$ FS $\gg$ 12}
\ElsIf{\texttt{FS < 0x100000}}
\State \texttt{R$\gets$ ((FS $\gg$ 12)*30)/100}
\ElsIf{\texttt{FS < 0xA00000}}
\State \texttt{R$\gets$ ((FS $\gg$ 12)*20)/100}
\ElsIf{\texttt{FS < 0x6400000}}
\State \texttt{R$\gets$ ((FS $\gg$ 12)*10)/100}
\ElsIf{\texttt{FS < 0x40000000}}
\State \texttt{R$\gets$ ((FS $\gg$ 12)*5)/100}
\Else
\State \texttt{R$\gets$ ((FS $\gg$ 12)*1)/100}
\EndIf
\If{\texttt{R == 1}}
    \Return \texttt{NBS}
\EndIf

\State \texttt{NBS $\gets$ (FS-(R $\ll$ 12))/(R-1)}
\State 
\Return \texttt{NBS}

\end{algorithmic}
\end{algorithm}

\begin{algorithm}[!ht]
\caption{Calculation of the start offsets of Keystream1 and Keystream2}\label{alg:calc_BS}
\begin{algorithmic}
\Require Infected file name
\Ensure Keystream offset SP1, SP2

\State \texttt{R $\gets$ base64urldecode(Filename)[16:]} 
\State \texttt{R1 $\gets$ byte to int64(R[:8])} \Comment {little endian}
\State \texttt{R2 $\gets$ byte to int64(R[8:])} \Comment {little endian}

\State \texttt{SP1 $\gets$ R1$\%$0x900000}

\State \texttt{SP2 $\gets$ R2$\%$0x9FFC00}
\end{algorithmic}
\end{algorithm}

\begin{algorithm}[!ht]
\caption{Hive ransomware master key recovery }\label{alg:calc_BS}
\begin{algorithmic} 
\Require Infected files, Original files
\Ensure Recovered Hive Ransomware master key

\State \Comment{Extract to equations using original file and infected file}
\For{\texttt{IF, OF in (Infected files, Original files)}}
    \State \texttt{NBS $\gets$ calc\_NBS(IF.size)} \Comment {Using algorithm 1}
    \State \texttt{SP1, SP2 $\gets$ calc\_offset(IF.name)} \Comment {Using algorithm 2}
    \State \texttt{iter  $\gets$ IF.size}/(0x1000+\texttt{NBS})
    
    \State \texttt{offset=0}
    \State \texttt{EQS $\gets$ set()}
    \For{\texttt{i $\gets$ 0 $\cdots$ iter}}
    \If{\texttt{i==iter}}
    \State \texttt{offset $\gets$ final encryption block offset}
    \Comment {See Section~\ref{hive_analysis}}
    \EndIf
        \For{\texttt{j $\gets$ 0 $\cdots$ 0xFFF}}
        
            \State \texttt{O1 $\gets$ offset\%0x100000}
            \State \texttt{O2 $\gets$ offset\%0x1000}
            
            \State \texttt{EQS.add(SP1+O1, SP2+O2, IF[offset]$\bigoplus$ OF[offset])}
            \State \Comment{IF[offset]$\bigoplus$ OF[offset] == byte of $EKS$}
             \State \texttt{offset+=1}
         \EndFor
\State \texttt{offset+=NBS}
\EndFor
\EndFor
\State \Comment{Extract equation end}

\State \texttt{EK $\gets$ [None]$\times$0xA00000}
\State \texttt{E $\xleftarrow[]{\$}$ EQS}
\State \texttt{EK[E[0]] $\xleftarrow[]{\$}$ \{0$\cdots$255\} }
\State \texttt{EQS $\gets$ tuple(EQS)}

 \algstore{part3}
\end{algorithmic}
\end{algorithm}
\begin{algorithm}[!ht]
  \begin{algorithmic}
    \algrestore{part3}

\While{Until the EQS size does not change}
\For{\texttt{EQ in EQS}}
\If{\texttt{(EK[EQ[0]] == None) and (EK[EQ[1]] == None)}}
\State \textbf{Continue}  
\ElsIf{\texttt{(EK[EQ[0]] != None) and (EK[EQ[1]] == None)}}
\State \texttt{EK[EQ[1]] = EK[EQ[0]] $\bigoplus$ EK[EQ[2]]}
\ElsIf{\texttt{(EK[EQ[0]] == None) and (EK[EQ[1]] != None)}}
\State \texttt{EK[EQ[0]] = EK[EQ[1]] $\bigoplus$ EK[EQ[2]]}
\ElsIf{\texttt{(EK[EQ[0]] != None) and (EK[EQ[1]] != None)}}
\State \texttt{EQS.pop(EQ)}
\EndIf
\EndFor
\EndWhile
\end{algorithmic}
\end{algorithm}

\newpage
\subsection{Experiments}

Experiments were conducted after infecting Windows 7, 10 (x86, x64) with the Hive ransomware. 
The experiments comprises two parts. The first experiment used a randomly created dataset, to determine the master key recovery rate.
In the second experiment, considering the actual environment, only data that could be downloaded online was searched and used without data backup.
First, we infected the experimental environment with Hive Ransomware, and then collected files that are guessed to be original files through the Internet, e-mail, and backup files.
We checked the unencrypted areas by XORing the collected and encrypted files. If XOR results are all 0x00, we confirmed the collected file as the original file and listed it.
If even one byte is not 0x00 in every area in the XOR result, it was never the original file.
However, it is possible that all areas are 0x00 but not the original file.
Therefore, if there is a contradiction in the process of solving the equation, we exclude the file from the original file list and solve the equation again.
We performed this process using a Proof-of-Concept code made in Python, and the master key was recovered through this.
The following is the first experimental result of an master key recovery experiment (Table.~\ref{tab:exp_result}).

\begin{table}[!ht]
\centering
\caption{Hive ransomware master key recovery experiment results using dataset}
\label{tab:exp_result}
\resizebox{0.5\textwidth}{!}{
\begin{tabular}{ccc} 
\hline
\multicolumn{1}{l}{Average file size}         & Number of files & Master key recovery rate (\%) \\ \hline
\multirow{4}{*}{\begin{tabular}[c]{@{}c@{}}21KB\\ ($\pm$ 5KB)\end{tabular}} & 500 & 61.69\\& 1,000& 81.53\\& 3,200& 92.68\\& 6,400& 94.59\\ \hline
\multirow{4}{*}{\begin{tabular}[c]{@{}c@{}}150KB\\ ($\pm$ 15KB)\end{tabular}} & 200 & 53.90\\& 300 & 69.42\\& 500 & 82.84\\& 1,000& 91.51\\ \hline
\multirow{5}{*}{\begin{tabular}[c]{@{}c@{}}501KB\\ ($\pm$ 5KB)\end{tabular}}  & 100 & 73.75\\& 200 & 88.04\\& 300 & 92.32\\& 500 & 93.75\\& 1,000& 95.03\\ \hline
\multirow{6}{*}{\begin{tabular}[c]{@{}c@{}}1MB\\ ($\pm$ 100KB)\end{tabular}} & 50  & 63.2\\& 100 & 88.54\\& 200 & 94.47\\& 300 & 95.59\\& 400 & 97.45\\& 500 & 98.05\\ \hline
\multirow{5}{*}{\begin{tabular}[c]{@{}c@{}}5MB\\ ($\pm$ 100KB)\end{tabular}}  & 30  & 74.5\\
& 50  & 88  \\& 100 & 95.21\\& 200 & 97.85\\& 300 & 99.04\\ \hline
\multirow{3}{*}{\begin{tabular}[c]{@{}c@{}}10MB\\ ($\pm$ 100KB)\end{tabular}} & 50  & 62.36\\& 100 & 87.1\\& 200 & 95.36    
\end{tabular}
}
\end{table}

\begin{figure}[ht!]
    \centering
    \includegraphics[width=0.9\textwidth]{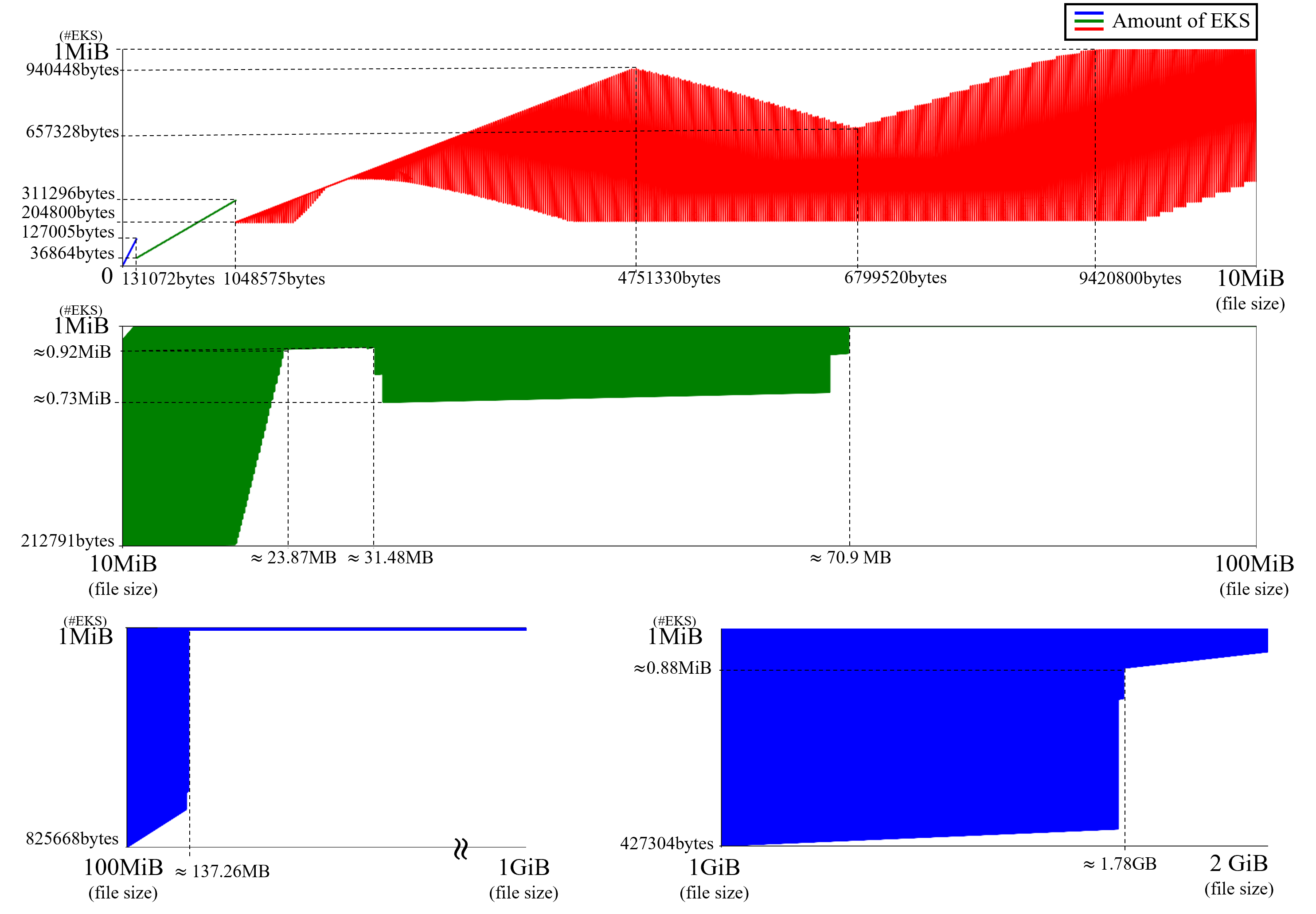}
    \caption{Amount of extractable EKS by file size}
    \label{fig:eks}
\end{figure}

The Hive ransomware master key recovery experiment required the use of a large number of files to recover more than 90\% of the key when the files were less than 500KB.
When we collected $EKS$ from files larger than 1MiB, we could recover more than 90\% of Hive master key with 100–200 files.
This difference in the number of files is because the number of encryption times, that is, the amount of $EKS$ that can be collected, varies according to the file size (cf. Table~\ref{tab:calculating_BS}).
The number of $EKS$ that can be collected varies depending on the $NBS$; the smaller the file size, the smaller the amount of $EKS$ that can be collected. 
Generally, the larger the file size, the larger the amount of $EKS$ we can collected, but in some cases the amount of the $EKS$ may decrease.
As a result of our experiment, it was possible to effectively acquire $EKS$ when using files larger than 0x280A000 bytes.
Figure~\ref{fig:eks} shows the available $EKS$ by file size.

We tried master key recovery by mixing various file sizes, in consideration of $NBS$, and found that more effective master key recovery was possible (Table.~\ref{tab:exp_result2}).

\begin{table}[!ht]
\centering
\caption{Result of the Hive ransomware master key recovery experiment with actual environment}
\label{tab:exp_result2}
\resizebox{0.6\textwidth}{!}{
\begin{tabular}{ccc}
\hline
Average file size & Number of files & Master key recovery rate (\%) \\ \hline
1KB & 9 & \multirow{5}{*}{95.85} \\
2–127KB & 24 &  \\
128–1,023KB & 19 &  \\
1–10MB & 19 &  \\
10–100MB & 20 &  \\ \hline
\end{tabular}
}
\end{table}

\begin{figure}[ht!]
    \centering
    \includegraphics[width=0.9\textwidth]{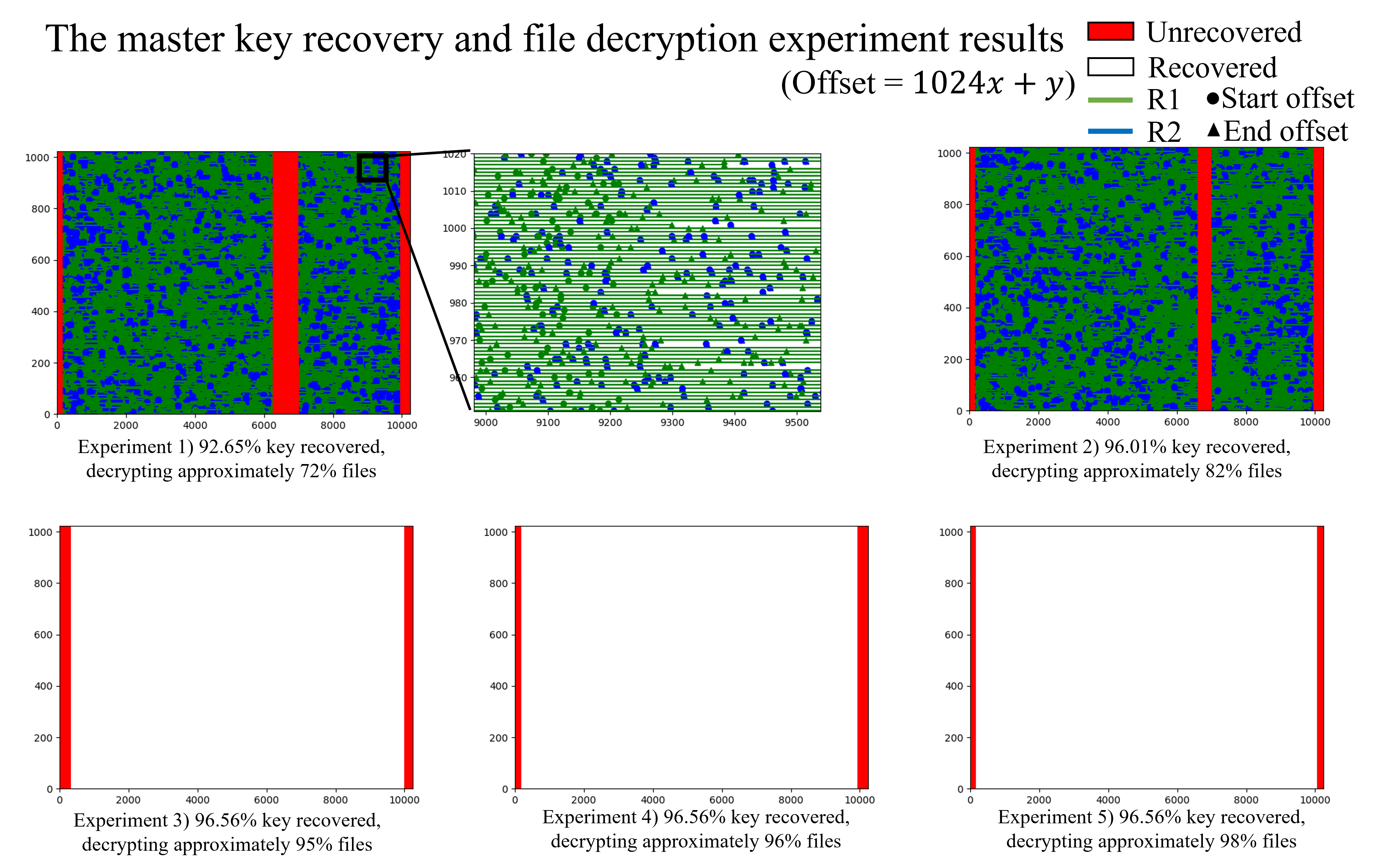}
    \caption{Visualization of the unrecovered master key area by experiment and the decrypted ratio of real data through the recovered master key}
    \label{fig:final}
\end{figure}

The next experiment is a decryption experiment of infected files using the recovered master key.
We infected 50,000 files and collected $EKS$ by using 50 files of 40MB size.
The master key was recovered through the collected $EKS$ and the data was decrypted using the master key.
We set the decryption success only when all bytes were decrypted, and when even a single byte could not be decrypted, it was set as a failure rather than a partial success.
In the case of partial success, additional collection of $EKS$ is possible and help to increase the recovery rate, but it was not considered in this experiment.
Figure~\ref{fig:final} is a visualization of the results of the master key recovery and file decryption experiment and the unrecovered area.

The experiment 1—2 shows both the unrecovered area of the master key and (R1, R2) used for file encryption.
The experiment 3—5 shows only the recovered and unrecovered areas of the master key except for (R1, R2).
As Hive ransomware uses secure random, (R1, R2) are evenly distributed throughout the master key offset.
Therefore, the recovery rate of the master key and the decryption rate of the entire file should be proportional.
However, in the experiment 1—2, there are many files that can be partially recovered, so the data recovery rate was much lower than in the experiment 3—5.

We found that the first and last parts of the master key were unrecovered in most cases.
To recover the first and last parts of the master key, the original file of the infected file with SP1 of 0x0, 0x900000 or close to it is required.
The results of recovering encrypted data using the recovered master key are as follows:
The master key recovered 92\% succeeded in decrypting approximately 72\% of the files, the master key restored 96\% succeeded in decrypting approximately 82\% of the files, and the master key restored 98\% succeeded in decrypting approximately 98\% of the files.

\section{Conclusions}
The Hive ransomware, which first appeared in June 2021, threatens individuals and enterprises, governments, critical infrastructure, and essential services. 
This ransomware forces sensitive data not only to be decrypted, but also to be leaked through HiveLeaks.
The Hive ransomware is so threatening that the FBI has released a warning report.

In this paper we presented a method for decrypting files encrypted with the Hive ransomware. 
The decryption method is feasible without access to the attacker’s information, using just encrypted files. 
We obtained the master key by solving numerous equations for XOR operations acquired from the encrypted files. 
We also experimentally verified that our method succeeded in the key recovery with approximately a rate of 95\%.
We expect that our method will be helpful for  individuals and enterprises damaged by the Hive ransomware.

\section*{Acknowledgement}
\label{ack}

This work was supported by Korea Information Security Agency (KISA) grant funded by the Korea government (Ministry of Science and ICT, MSIT)
\bibliographystyle{elsarticle-num}
\bibliography{bibliography}

\begin{thebibliography}{10}
\expandafter\ifx\csname url\endcsname\relax
  \def\url#1{\texttt{#1}}\fi
\expandafter\ifx\csname urlprefix\endcsname\relax\def\urlprefix{URL }\fi
\expandafter\ifx\csname href\endcsname\relax
  \def\href#1#2{#2} \def\path#1{#1}\fi

\bibitem{mouratidis2021covid}
K.~Mouratidis, A.~Papagiannakis, Covid-19, internet, and mobility: The rise of
  telework, telehealth, e-learning, and e-shopping, Sustainable Cities and
  Society 74 (2021) 103182.

\bibitem{SecurityMagazine}
{2020 Vulnerability and Threat Trends Report},
  \url{https://lp.skyboxsecurity.com/WICD-2020-07-WW-VT-Trends_Asset.html},
  accessed: 20-Dec-2021.

\bibitem{richardson2017ransomware}
R.~Richardson, M.~M. North, Ransomware: Evolution, mitigation and prevention,
  International Management Review 13~(1) (2017) 10.

\bibitem{Hopitalde}
{Hôpital de Dax totalement bloqué : Twitter, ultime recours face au
  ransomware},
  \url{https://www.zdnet.fr/actualites/hopital-de-dax-totalement-bloque-twitter-ultime-recours-face-au-ransomware-39917771.htm},
  accessed: 20-Dec-2021.

\bibitem{UnitingCareQueensland}
{UnitingCare Queensland hit by cyber attack},
  \url{https://www.computerweekly.com/news/252499835/UnitingCare-Queensland-hit-by-cyber-attack},
  accessed: 20-Dec-2021.

\bibitem{ScrippsHealth}
{Health care giant Scripps Health hit by ransomware attack},
  \url{https://www.bleepingcomputer.com/news/security/health-care-giant-scripps-health-hit-by-ransomware-attack/},
  accessed: 20-Dec-2021.

\bibitem{HSE}
{Ireland's Health Services hit with \$20 million ransomware demand},
  \url{https://www.bleepingcomputer.com/news/security/irelands-health-services-hit-with-20-million-ransomware-demand/},
  accessed: 20-Dec-2021.

\bibitem{Dutchsupermarkets}
{Dutch supermarkets run out of cheese after ransomware attack},
  \url{https://www.bleepingcomputer.com/news/security/dutch-supermarkets-run-out-of-cheese-after-ransomware-attack/},
  accessed: 20-Dec-2021.

\bibitem{JBS}
{JBS paid \$11 million to REvil ransomware, \$22.5M first demanded},
  \url{https://www.bleepingcomputer.com/news/security/jbs-paid-11-million-to-revil-ransomware-225m-first-demanded/},
  accessed: 20-Dec-2021.

\bibitem{Colonial}
{Colonial Pipeline paid \$5 million ransom to hackers},
  \url{https://www.cnbc.com/2021/05/13/colonial-pipeline-paid-ransom-to-hackers-source-says.html},
  accessed: 20-Dec-2021.

\bibitem{homeaffairs}
{RANSOMWARE ACTION PLAN},
  \url{https://www.homeaffairs.gov.au/cyber-security-subsite/files/ransomware-action-plan.pdf},
  accessed: 20-Dec-2021.

\bibitem{Canadian}
{Ransomware: How to prevent and recover (ITSAP.00.099)},
  \url{https://cyber.gc.ca/en/guidance/ransomware-how-prevent-and-recover-itsap00099},
  accessed: 20-Dec-2021.

\bibitem{weekend_alert}
{Ransomware Awareness for Holidays and Weekends},
  \url{https://us-cert.cisa.gov/ncas/alerts/aa21-243a}, accessed: 20-Dec-2021.

\bibitem{FBI_Hive}
{Indicators of Compromise Associated with Hive Ransomware},
  \url{https://www.documentcloud.org/documents/21049431-fbi-flash-hiveꠓransomware-iocs},
  accessed: 20-Dec-2021.

\bibitem{hivebppage}
{Hive ransomware enters big league with hundreds breached in four months},
  \url{https://www.bleepingcomputer.com/news/security/hive-ransomware-enters-big-league-with-hundreds-breached-in-four-months/},
  accessed: 20-Dec-2021.

\bibitem{Altus}
{New ransomware group Hive leaks Altus group sample files},
  \url{https://cybernews.com/news/new-ransomware-group-hive-leaks-altus-group-sample-files/},
  accessed: 20-Dec-2021.

\bibitem{Memorial}
{Hive ransomware attacks Memorial Health System, steals patient data},
  \url{https://www.bleepingcomputer.com/news/security/hive-ransomware-attacks-memorial-health-system-steals-patient-data/},
  accessed: 20-Dec-2021.

\bibitem{Missouri}
{Hive ransomware group attacks Missouri health center},
  \url{https://www.healthcareitnews.com/news/hive-ransomware-group-attacks-missouri-health-center},
  accessed: 20-Dec-2021.

\bibitem{Macquarie}
{Macquarie Health Corporation hit by Windows Hive ransomware},
  \url{https://itwire.com/security/macquarie-health-corporation-hit-by-windows-hive-ransomware.html},
  accessed: 20-Dec-2021.

\bibitem{MediaMarkt}
{MediaMarkt hit by Hive ransomware, ransom now at 50 million},
  \url{https://news.securiwiser.com/mediamarkt-hit-by-hive-ransomware-ransom-now-at-50-million/},
  accessed: 20-Dec-2021.

\bibitem{rfc3826}
F.~Maino, U.~Blumenthal, K.~McCloghrie,
  \href{https://www.rfc-editor.org/info/rfc3826}{{The Advanced Encryption
  Standard (AES) Cipher Algorithm in the SNMP User-based Security Model}}, RFC
  3826 (Jun. 2004).
\newblock \href {https://doi.org/10.17487/RFC3826}
  {\path{doi:10.17487/RFC3826}}.
\newline\urlprefix\url{https://www.rfc-editor.org/info/rfc3826}

\bibitem{rfc8017}
K.~Moriarty, B.~Kaliski, J.~Jonsson, A.~Rusch,
  \href{https://www.rfc-editor.org/info/rfc8017}{{PKCS \#1: RSA Cryptography
  Specifications Version 2.2}}, RFC 8017 (Nov. 2016).
\newblock \href {https://doi.org/10.17487/RFC8017}
  {\path{doi:10.17487/RFC8017}}.
\newline\urlprefix\url{https://www.rfc-editor.org/info/rfc8017}

\bibitem{kharraz2015cutting}
A.~Kharraz, W.~Robertson, D.~Balzarotti, L.~Bilge, E.~Kirda, Cutting the
  gordian knot: A look under the hood of ransomware attacks, in: International
  Conference on Detection of Intrusions and Malware, and Vulnerability
  Assessment, Springer, 2015, pp. 3--24.

\bibitem{cabaj2018software}
K.~Cabaj, M.~Gregorczyk, W.~Mazurczyk, Software-defined networking-based crypto
  ransomware detection using http traffic characteristics, Computers \&
  Electrical Engineering 66 (2018) 353--368.

\bibitem{scaife2016cryptolock}
N.~Scaife, H.~Carter, P.~Traynor, K.~R. Butler, Cryptolock (and drop it):
  stopping ransomware attacks on user data, in: 2016 IEEE 36th International
  Conference on Distributed Computing Systems (ICDCS), IEEE, 2016, pp.
  303--312.

\bibitem{chen2017automatic}
Z.-G. Chen, H.-S. Kang, S.-N. Yin, S.-R. Kim, Automatic ransomware detection
  and analysis based on dynamic api calls flow graph, in: Proceedings of the
  International Conference on Research in Adaptive and Convergent Systems,
  2017, pp. 196--201.

\bibitem{alhawi2018leveraging}
O.~M. Alhawi, J.~Baldwin, A.~Dehghantanha, Leveraging machine learning
  techniques for windows ransomware network traffic detection, in: Cyber threat
  intelligence, Springer, 2018, pp. 93--106.

\bibitem{mehnaz2018rwguard}
S.~Mehnaz, A.~Mudgerikar, E.~Bertino, Rwguard: A real-time detection system
  against cryptographic ransomware, in: International Symposium on Research in
  Attacks, Intrusions, and Defenses, Springer, 2018, pp. 114--136.

\bibitem{kok2020evaluation}
S.~Kok, A.~Azween, N.~Jhanjhi, Evaluation metric for crypto-ransomware
  detection using machine learning, Journal of Information Security and
  Applications 55 (2020) 102646.

\bibitem{continella2016shieldfs}
A.~Continella, A.~Guagnelli, G.~Zingaro, G.~De~Pasquale, A.~Barenghi,
  S.~Zanero, F.~Maggi, Shieldfs: a self-healing, ransomware-aware filesystem,
  in: Proceedings of the 32nd Annual Conference on Computer Security
  Applications, 2016, pp. 336--347.

\bibitem{jung2018ransomware}
S.~Jung, Y.~Won, Ransomware detection method based on context-aware entropy
  analysis, Soft Computing 22~(20) (2018) 6731--6740.

\bibitem{cabaj2016using}
K.~Cabaj, W.~Mazurczyk, Using software-defined networking for ransomware
  mitigation: the case of cryptowall, Ieee Network 30~(6) (2016) 14--20.

\bibitem{kharaz2016unveil}
A.~Kharaz, S.~Arshad, C.~Mulliner, W.~Robertson, E.~Kirda, $\{$UNVEIL$\}$: A
  large-scale, automated approach to detecting ransomware, in: 25th
  $\{$USENIX$\}$ Security Symposium ($\{$USENIX$\}$ Security 16), 2016, pp.
  757--772.

\bibitem{xu2017cryptographic}
D.~Xu, J.~Ming, D.~Wu, Cryptographic function detection in obfuscated binaries
  via bit-precise symbolic loop mapping, in: 2017 IEEE Symposium on Security
  and Privacy (SP), IEEE, 2017, pp. 921--937.

\bibitem{li2018k}
J.~Li, Z.~Lin, J.~Caballero, Y.~Zhang, D.~Gu, K-hunt: Pinpointing insecure
  cryptographic keys from execution traces, in: Proceedings of the 2018 ACM
  SIGSAC Conference on Computer and Communications Security, 2018, pp.
  412--425.

\bibitem{hill2018cryptoknight}
G.~Hill, X.~Bellekens, Cryptoknight: generating and modelling compiled
  cryptographic primitives, Information 9~(9) (2018) 231.

\bibitem{villalba2018ransomware}
L.~J.~G. Villalba, A.~L.~S. Orozco, A.~L. Vivar, E.~A.~A. Vega, T.-H. Kim,
  Ransomware automatic data acquisition tool, IEEE Access 6 (2018)
  55043--55052.

\bibitem{lee2021magniber}
S.~Lee, M.~Park, J.~Kim, Magniber v2 ransomware decryption: Exploiting the
  vulnerability of a self-developed pseudo random number generator, Electronics
  10~(1) (2021) 16.

\bibitem{rukhin2001nist}
A.~Rukhin, Nist special publication 800-22, http://csrc. nist.
  gov/rng/SP800-22b. pdf (2001).

\bibitem{lee2019study}
S.~Lee, B.~Youn, S.~Kim, G.~Kim, Y.~Lee, D.~Kim, H.~Park, J.~Kim, A study on
  encryption process and decryption of ransomware in 2019, Journal of the Korea
  Institute of Information Security \& Cryptology 29~(6) (2019) 1339--1350.

\bibitem{lee2019machine}
K.~Lee, S.-Y. Lee, K.~Yim, Machine learning based file entropy analysis for
  ransomware detection in backup systems, IEEE Access 7 (2019) 110205--110215.

\bibitem{kolodenker2017paybreak}
E.~Kolodenker, W.~Koch, G.~Stringhini, M.~Egele, Paybreak: Defense against
  cryptographic ransomware, in: Proceedings of the 2017 ACM on Asia Conference
  on Computer and Communications Security, 2017, pp. 599--611.

\bibitem{cheng2019dptcry}
G.~Cheng, C.~Guo, Y.~Tang, dptcry: an approach to decrypting ransomware
  wannacry based on api hooking, CCF Transactions on Networking 2~(3) (2019)
  207--216.

\end{thebibliography}

\end{document}